\documentclass[pra,twocolumn,amssymb,superscriptaddress]{revtex4-2}

\usepackage[breaklinks=true,
colorlinks=true,
pdfusetitle=true]{hyperref}
\usepackage{physics}
\usepackage{qcircuit}
\usepackage[ruled,vlined,linesnumbered]{algorithm2e}
\usepackage{calc}
\usepackage{graphicx}

\usepackage{amsthm}
\newtheorem{theorem}{Theorem}
\newtheorem{lemma}[theorem]{Lemma}

\DeclareMathOperator*{\dg}{\dagger}
\newcommand{\aqpe}{\texorpdfstring{\(\alpha\)}{alpha}-QPE}
\newcommand{\bigO}{\mathcal{O}}
\newcommand{\good}{\ket{\text{good}}}
\newcommand{\bad}{\ket{\text{bad}}}

\def\et al.{\/\textit{et al.}}

\newcommand{\naturals}{\mathbb{N}}
\newcommand{\reals}{\mathbb{R}}

\begin{document}

\title{Simplifying a classical-quantum algorithm interpolation \\ with quantum singular value transformations}

\author{Duarte \surname{Magano}}
\thanks{These authors contributed equally to the work}
\email[corresponding author address: ]{miguel.murca@tecnico.ulisboa.pt}

\author{Miguel \surname{Mur\c{c}a}}
\thanks{These authors contributed equally to the work}
\email[corresponding author address: ]{miguel.murca@tecnico.ulisboa.pt}

\affiliation{Instituto Superior T\'{e}cnico, Universidade de Lisboa, Portugal}
\affiliation{Instituto de Telecomunica\c{c}\~{o}es, Lisboa, Portugal}

\date{\today}

\begin{abstract}
    The problem of Phase Estimation (or Amplitude Estimation) admits a quadratic
    quantum speedup.  Wang, Higgott and Brierley [2019, Phys. Rev. Lett. 122
            140504] have shown that there is a continuous trade-off between quantum
    speedup and circuit depth (by defining a family of algorithms known as
    \aqpe).
    In this work, we show that the scaling of \aqpe\ can be naturally and
    succinctly derived within the framework of Quantum Singular Value
    Transformation (QSVT).
    From the QSVT perspective, a greater number of coherent oracle calls translates
    into a better polynomial approximation to the sign function, which is the
    key routine for solving Phase Estimation.  The better the approximation to
    the sign function, the fewer samples one needs to determine the sign
    accurately.
    With this idea, we simplify the proof of \aqpe, while providing a new
    interpretation of the interpolation parameters, and show that QSVT is a
    promising framework for reasoning about classical-quantum interpolations.
\end{abstract}

\maketitle

\section{Introduction}

{
With the developments in practical implementations of quantum computers, and
with fault-tolerant computation still apparently beyond reach in the near
future, much attention has been focused on classical-quantum (hybrid)
algorithms: those that can leverage a limited amount of quantum coherence, while
out-performing completely classical algorithms \cite{Sun2019,Weigold2021,Callison2022,Arora22}. 
That such hybrid algorithms exist for any given problem, and for any coherence time constraint, is not obvious. 
Explicitly showing the existence of a continuous trade-off between classical
and quantum resources provides much insight on the role of
the two types of resources in the algorithm, and how to off-load work from one
to the other. Of course, the formulation of hybrid algorithms gives also the
practical advantage of being able to adapt to the available quantum resources,
such that, even when these are limited, a speed-up can be achieved.

One particular problem for which hybrid algorithms have been thoroughly studied is
that of Phase Estimation. 
In this problem, we are given access to a unitary $U$ and an eigenstate $\ket{\psi}$ with eigenvalue $e^{i \phi}$, where the value of $\phi$ is unknown; the goal is to determine $\phi$.
To understand the hybrid algorithm approaches to Phase Estimation, it is useful to first consider the
Iterative Phase Estimation algorithm (also known as Kitaev's Phase
Estimation) \cite{Kitaev1995,Griffiths1996}. The method is obtained from the
usual Quantum Fourier Transform-based Phase Estimation by invoking the deferred measurement principle, which
produces the circuit of Figure~\ref{fig:circuit}.
The circuit has the free parameters $M$ and $\theta$, which are updated at each iteration of the algorithm.
Concretely, to obtain \(\phi\) up to a precision of \(2^{-m}\),
one should set, at each step,
\setlength{\dimen255}{\widthof{\((M, \theta)_3 = (2^{m-3}, -(2^{-m}\cdot\phi_m + 2^{-m+1}\cdot\phi_{m-1}))\)} / 2}
\begin{align*}
     & (M, \theta)_1 = (2^{m-1}, 0)                                                      \\
     & (M, \theta)_2 = (2^{m-2}, -\pi\cdot2^{-m}\cdot\phi_m)                             \\
     & (M, \theta)_3 = (2^{m-3}, -\pi\cdot(2^{-m}\cdot\phi_m + 2^{-m+1}\cdot\phi_{m-1})) \\
     & \hspace*{\the\dimen255} \vdots                                                    \\
     & (M, \theta)_m = (1, -\pi\cdot(2^{-m}\cdot\phi_m + \cdots + 2^{-2}\cdot\phi_{2}))
\end{align*}
where, at step \(j\), if \(E\) is measured to be \(1\), then \(\phi_{m-j+1} = 1\), and
if \(E = 0\), \(\phi_{m-j+1} = 0\). By the end of the procedure, \(\phi \approx \pi
\sum_{j=1}^{m} 2^{-j} \phi_j\), with precision \(2^{-m}\).

\begin{figure}[hbt]
    \includegraphics[width=\linewidth]{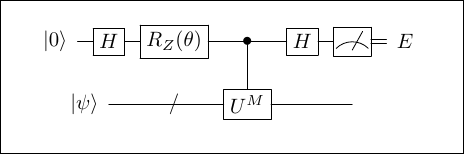}
    \caption{Circuit for Iterative Phase Estimation. \(U \ket{\psi} = e^{i \phi} \ket{\psi}\),
    \(R_Z(x) = \dyad{0} + e^{ix} \dyad{1}\), \(M \in \naturals\), \(\theta \in \reals\), and
    $E \in \{0,1\}$, with \(M, \theta\) to be set as part of the algorithm.}
    \label{fig:circuit}
\end{figure}

This formulation opens the possibility for other choices of the update method for \(M\) and \(\theta\).
Svore \et al.\ \cite{Svore2013} note that an
informational perspective can be adopted, whereby one actually wishes to
estimate a parameter of a distribution, and should seek to maximize the Fisher
information of their measurements. By exploiting schedules that maximize
information, they reduced the number of necessary samples in logarithmic factors, and approach that they call Fast Phase Estimation.

Nonetheless, both Fast Phase Estimation and the Iterative Phase Estimation have, up to polylogarithmic
factors, the same requirements in terms of the total number of calls to $U$, denoted by \(T\), and
the required circuit depths, denoted by \(D\): \(T = \tilde\bigO(1/\epsilon), D =
\bigO(1/\epsilon)\), where \(\epsilon\) is the precision error attained in the
estimation of \(\phi\) \footnote{We adopt the standard ``big O'' notation for asymptotic upper bounds. For two functions $f$ and $g$ from $\mathbb{R}$ to $\mathbb{R}$ we say that $f=\bigO(g)$ if $\exists C, x_0 > 0: \forall x, \left(x > x_0 \implies f(x) < C \cdot g(x) \right)$. The $\tilde\bigO$ means that we ignore poly-logarithmic terms. That is, $\bigO(f(x) \log^c x) = \tilde\bigO(f(x))$.}. 
These requirements should be contrasted with those for
the Hadamard Test, which corresponds essentially to a classical statistical
sampling approach: the quantity of interest is encoded in the odds of a
Bernoulli distribution, which can be optimally estimated to precision
\(\epsilon\) with \(\bigO(1/\epsilon^2)\) samples of the circuit
\cite{Cramer1946}, but with a single oracle call per run (thus with \(D =
\bigO(1)\)).

Explorations of alternative schedules and associated classical algorithms
brought Wiebe \et al.\ to formulate Bayesian Phase Estimation \cite{Wiebe2016},
eventually leading to Wang \et al.'s \(\alpha\)-Quantum Phase Estimation (\aqpe) \cite{Wang2019}. This
approach established a continuous trade-off between depth and sampling
complexity, and bridged the aforementioned Iterative Phase Estimation and statistical sampling.
Giurgica-Tiron \et al.\ \cite{GiurgicaTiron2022} went on to rigorously show the
convergence of these methods, and connected these hybridized Phase Estimation
algorithms with other results on Quantum Fourier Transform-free Amplitude Estimation algorithms
\cite{Aaronson2020,Suzuki2020}.

Notably, \aqpe\ describes the spectrum of hybrid Phase Estimation algorithms
with a single scalar parameter --- the titular \(\alpha\) --- such that the
depth and sample complexity attained for a given choice of \(\alpha\) and
precision \(\epsilon\) are \(T = \bigO(1/{\epsilon^{1+\alpha}})\), and \(D =
\bigO(1/{\epsilon^{1-\alpha}})\), respectively. Note how the trade-off \(TD\)
remains a constant \(\bigO(1/{\epsilon^2})\), and how this relationship is also
verified both for statistical sampling and Iterative Phase Estimation.

Furthermore, Phase Estimation enjoys a close relationship with other notable algorithms;
Brassard \et al.'s algorithm for Amplitude Estimation \cite{Brassard2002}
relies on it for a Grover-like operator. This places both
Amplitude Estimation and Phase Estimation as central ingredients in many of quantum computing's celebrated
applications, such as Quantum Counting \cite{Brassard98}, Quantum Montecarlo
\cite{Montanaro2015}, Quantum Linear Systems \cite{Harrow2009} or Ground State
Preparation \cite{Abrams99}. One also finds that using this relationship and the
machinery of \aqpe, a classical-quantum interpolation for Amplitude Estimation can likewise be
found.
As we will see, a similar relation also holds for the problem of Eigenvalue Estimation.

Recently, Gily\'{e}n \et al.\ introduced Quantum Singular Value Transformation
(QSVT) \cite{Gilyen2019,Martyn2021}. As a generalization of the work of Quantum
Signal Processing \cite{Low2019}, it has proven to be an extremely powerful
framework for describing quantum computation, having been shown by its authors
to successfully describe quantum algorithms for search, Phase Estimation, and
various quantum linear algebra results, among other applications.

In this work, we show that \aqpe\ follows naturally from a Quantum Singular Value Transformation construction for Eigenvalue Estimation.
This greatly simplifies the derivation of \aqpe, if one is familiar with the
main results of QSVT.  At the same time, our method provides a different
interpretation of the scalar parameter \(\alpha\), namely relating it to the
precision with which a step function is approximated by a constrained
polynomial. Finally, this work may serve as a starting point for hybridizations
of other relevant high-coherence algorithms under the QSVT description.
}

\section{Preliminaries}

\begin{figure*}
    \hskip -.5in 
    \includegraphics[width=1\linewidth]{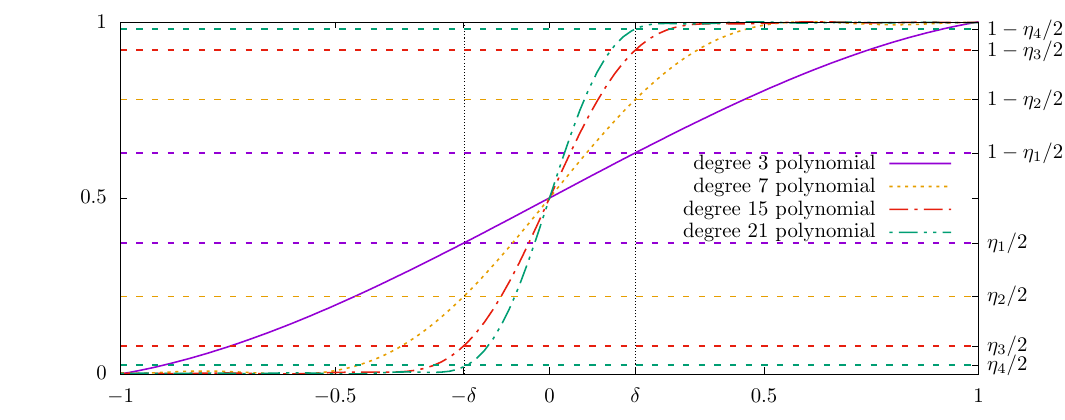}
    \caption{
        For a fixed \(\delta\) (here, \(\delta = 0.2\)), polynomials of
        increasing degree allow for a better approximation to the step function
        (equivalently, with a smaller \(\eta\)). While this decreases the
        necessary number of samples, the implementation of polynomials of higher
        degrees in the Quantum Singular Value Transformation framework requires
        longer coherent computation. A theorem of \cite{Lin2020} ensures the
        existence of a suitable polynomial for a given choice of \(\delta,\eta\)
        (cf. lemma \ref{thm:projector}); here we used the \texttt{pyqsp}
        \cite{Martyn2021,Chao2020,pyqsp} package to explicitly generate such polynomials.}
    \label{fig:polynomial}
\end{figure*}

{\subsection{Computational model}

The analysis of \aqpe\ is naturally set in a hybrid computing model
\cite{Sun2019,Arora22}. A hybrid classical-quantum algorithm is one that performs
multiple runs of limited-depth quantum circuits, possibly interchangeably with
some classical processing of the measurement outcomes.

The problems that we are concerned with here are defined in an oracular setting.
That is, they are specified in terms of access to some given operators that
encode the relevant information. Thus, our complexity measures are based on the
number of oracle calls (or \emph{queries}). The \emph{depth} complexity of the
algorithm, $D$, is the maximum depth among all the employed quantum circuits.
The \emph{time} complexity, $T$, is the total running time, that is, the sum of
all the depths or, equivalently, the total number of queries.

Evidently, every $T$-time, $D$-depth algorithm can be converted into a $T$-time,
$\bigO(T)$-depth algorithm \cite{Nielsen2010}. But, in the context of
limited-coherence computing, it becomes relevant to bound $D$ as much as
possible. Moreover, in certain situations it may be beneficial to lower the
depth complexity even if at the cost of increasing the total running time. This
trade-off between $T$ and $D$ is achieved by \aqpe.

\begin{theorem}[\aqpe, Proposition~1 of \cite{Wang2019} and Theorem~2.3 of \cite{GiurgicaTiron2022}]
    For any $\alpha \in [0,1]$, there is a $T(\alpha)$-time, $D(\alpha)$-depth
    algorithm that solves PE, where
    \begin{equation}
        T(\alpha) = \bigO\left( \frac{1}{\epsilon^{1 + \alpha}} \right)
        \; \text{and } \;
        D(\alpha) = \bigO\left( \frac{1}{\epsilon^{1 - \alpha}} \right).
    \end{equation}
    \label{thm:alphaQPE}
\end{theorem}

\subsection{Problem reductions \label{sec:reductions}}

As we have discussed, the problems of Phase Estimation and Amplitude Estimation are closely related. They also
relate to the problem of Eigenvalue Estimation. Below, we precisely define
each of these.

\textbf{Phase estimation (PE).}
Let $U$ be a unitary operator and $\ket{\psi}$ a state such that $U \ket{\psi} = e^{i \phi} \ket{\psi}$
for some unknown $\phi \in [0, 2 \pi)$. Let $U_{\psi}$ be an operator that
prepares $\ket{\psi}$: $U_{\psi} \ket{0^m} = \ket{\psi}$.
\textit{Input:} Access operators $U_{\psi}$, $U_{\psi}^{\dg}$, controlled-$U$,
and controlled-$U^{\dg}$, and a precision parameter $\epsilon > 0$.
\textit{Output:} An estimation of $\phi$ up to $\epsilon$, with bounded-error
probability.

\textbf{Amplitude estimation (AE).}
Let $A$ be a unitary operator such that $A \ket{0^m} = \sqrt{p} \good + \sqrt{1 - p^2} \bad$
and let $O_A$ be an oracle that distinguishes $\ket{\text{good}}$
from $\ket{\text{bad}}$ (say, by applying a $-1$ phase to $\ket{\text{good}}$).
\textit{Input:} Access to operators $A$, $A^{\dg}$, and $O_A$, and a precision
parameter $\epsilon > 0$.
\textit{Output:} An estimation of $\vert p \vert$ up to $\epsilon$, with
bounded-error probability.

Before defining Eigenvalue Estimation, we need to introduce the concept of block-encoding, which
permits representing non-unitary matrices in quantum circuits. We say that an
$n$-qubit matrix $H$ is $(\gamma, m)$--block-encoded in an $(m + n)$ unitary
matrix $U_H$ if
\begin{equation}
    H = \gamma \left( \bra{0^m} \otimes I_n \right) U_H \left( \ket{0^m} \otimes I_n \right).
\end{equation}

\textbf{Eigenvalue estimation (EE).}
Let $H$ be a Hermitian operator and $\ket{\psi}$ a state such that $H \ket{\psi}
    = \mu \ket{\psi} $ for some unknown $\mu$. Let $U_H$ be a $(\gamma,
    m)$--block-encoding of $H$ and let $U_{\psi}$ prepare $\ket{\psi}$: $U_{\psi}
    \ket{0^m} = \ket{\psi}$.
\textit{Input:} Access to operators $U_{\psi}$, $U_H$, and $U_H^{\dg}$, the
factor $\gamma$, and a precision parameter $\epsilon > 0$.
\textit{Output:} An estimation of $\mu$ up to $\epsilon$, with bounded-error
probability.

Typically, these three computational problems are taken to be equivalent.
This notion can be made rigorous. Given two problems
$\mathrm{Pr}_1$ and $\mathrm{Pr}_2$, we write $\mathrm{Pr}_1 \preceq
    \mathrm{Pr}_2$ if a $T$-time, $D$-depth algorithm to solve $\mathrm{Pr}_2$ can
be converted into a $\bigO(T)$-time, $\bigO(D)$-depth algorithm to solve
$\mathrm{Pr}_1$.
In Appendix \ref{ap:reductions}, we show the following.
\begin{lemma}
    $\mathrm{PE} \preceq \mathrm{AE} \preceq \mathrm{EE}$.
    \label{thm:reductions}
\end{lemma}

Throughout the rest of the article, we will only be concerned with the EE
problem. By Lemma \ref{thm:reductions}, our results apply immediately also to
PE and AE.

\subsection{Quantum eigenvalue transformations}

We approach EE with the filtering method developed by Lin and
Tong~\cite{Lin2020}, relying on the general theory for quantum singular value
transformations \cite{Gilyen2019}. Here we briefly review the main results that
we need.

Let $H$ be a Hermitian matrix with a spectral decomposition $\{\mu_i,
    \ket{\psi_i} \}_i$. For any function $F$, we define the eigenvalue
transformation $F(H)$ as
\begin{equation}
    H = \sum_i \mu_i \dyad{\psi_i} \rightarrow F(H) := \sum_i F(\mu_i) \dyad{\psi_i}.
\end{equation}
The idea of quantum eigenvalue transformations \footnote{The work of Gily\'{e}n \et al.\ \cite{Gilyen2019} describes a more
    general type of transformations, not restricted to Hermitian (square) matrices.
    We devote our attention to eigenvalue transformations, as opposed to the general
    singular value transformations, as those are the only ones that we use in this
    work.}\ is that, given a
block encoding of $H$, we can perform a very broad class of polynomial
transformations on $H$ in a time proportional to the degree of the polynomials.

\begin{theorem}[Quantum eigenvalue transformations, Theorem~2 of \cite{Gilyen2019}]
    Let $U_H$ be a $(\gamma, m)$--block-encoding of a Hermitian matrix $H$ and
    let $P \in \reals[x]$ be a $d$-degree polynomial with definite parity and
    $\vert P(x) \vert \leq 1$ for any $x \in [-1,1]$. Then, there is a $(1,m +
        1)$--block-encoding of $P(H / \gamma)$ using $d$ queries of $U_H$ and
    $U_H^{\dg}$.
    \label{thm:qet}
\end{theorem}

In particular, it is possible to approximate the sign function up to a desired
accuracy \cite[Lemma~14]{Gilyen2019}. This is used by Lin and
Tong~\cite{Lin2020} to block-encode an approximation of a projector onto the
subspace of eigenstates with eigenvalues larger than some threshold $\mu_0$.
See Figure \ref{fig:polynomial} for an illustration of this construction.

\begin{lemma}[Block-encoding approximation of step function, Lemma~5 of \cite{Lin2020}]
    Let $U_H$ be a $(\gamma, m)$-block-encoding of a Hermitian matrix $H$ and $\mu_0 \in [0, \gamma]$.
    Then, there is a $(1,m + 3)$--block-encoding of $P\left(\frac{H - \mu_0 I}{\gamma + \mu_0}; \delta , \eta \right)$, where $P$ satisfies
    \begin{align}
                      & \forall x \in [-1, -\delta], 0 \leq P(x; \delta, \eta) \leq \eta/2      \\
        \text{and }\, & \forall x \in [\delta, 1], 1 - \eta / 2 \leq P(x; \delta, \eta) \leq 1,
    \end{align}
    using $\bigO\left(\frac{1}{\delta} \log(\frac{1}{\eta})\right)$ queries of $U_H$ and $U_H^{\dg}$.
    \label{thm:projector}
\end{lemma}

In their work, Lin and Tong~\cite{Lin2020} apply this construction to the
problem of ground energy estimation with a binary search scheme. We adopt a
similar strategy to the EE problem, re-deriving the scaling of \aqpe.}

\section{\texorpdfstring{$\alpha$}{alpha}-QPE from quantum eigenvalue transformations}

{
\newcommand{\LEFT}{\texttt{LEFT}}
\newcommand{\RIGHT}{\texttt{RIGHT}}

The classical-quantum interpolation is perhaps simpler to appreciate for the
decision version of EE: given the same setting as EE and a parameter $\mu_0 \in
    [-\gamma, \gamma]$, the task is to determine with bounded-error probability
if $\mu$ is smaller than $\mu_0 - \epsilon / 2$ or greater than $\mu_0 +
    \epsilon / 2$, under the promise that one must be true. We focus on this problem
for now and later we see how to turn this into a solution for the complete
estimation task.

Using the construction from Lemma \ref{thm:projector}, we block-encode the
step function approximation $P$, centered at $\mu_0$. We then measure the first $m +
    3$ qubits (i.e., the block-encoding register), assigning an outcome
\RIGHT\ to $\ket{0^{m+3}}$ and an outcome \LEFT\ otherwise. Now,
choose
\begin{equation}
    \delta = \frac{\epsilon}{4 \gamma}.
\end{equation}
Then, if $\mu < \mu_0 - \epsilon / 2$, the probability of observing \RIGHT\ is
smaller than $(\eta / 2)^2$. In contrast, if $\mu > \mu_0 + \epsilon / 2$, the
probability of that outcome is greater than $(1 - \eta / 2)^2$. So, all that we
have to do is to distinguish the bias of the Bernoulli distribution of
\LEFT/\RIGHT\ outcomes with a precision smaller than
\begin{equation}
    \frac{1}{2} \left( (1 - \eta / 2)^2 - (\eta / 2)^2 \right) = \frac{1 - \eta}{2}.
\end{equation}
By Chebychev's inequality, we reach such an estimate with bounded-error
probability by taking
\begin{equation}
    \bigO\left( \frac{1}{(1 - \eta)^2}\right)
    \label{eq:numberoftrials}
\end{equation}
trials.

We have the freedom to tune $\eta$ as desired. The lower the value of $\eta$,
the fewer trials are necessary. On the other hand, a low $\eta$ requires a
polynomial of a high degree, meaning more coherent applications of $U_H$ and
$U_H^{\dg}$.

For example, with a single application of $U_H$ we can only prepare a polynomial
of degree one. In particular, the construction of Lemma \ref{thm:projector}
implements a block-encoding of $\frac{1}{2} \left(I +  \frac{H - \mu_0 I}{\gamma
        + \mu_0} \right)$. A straightforward calculation shows that, in this case, we
need to estimate the bias of the \LEFT/\RIGHT\ Bernoulli distribution with precision
at least $\epsilon / 4 \gamma$, requiring $\bigO( \gamma^2 / \epsilon^2)$
trials. This is but a classical statistical sampling approach.

Considering the other extreme case, say that we just want to take $\bigO(1)$
trials. From expression \eqref{eq:numberoftrials}, we then need to set $\eta = 1 -
    \bigO(1)$, which leads to circuits of depth
\begin{equation}
    \bigO\left(\frac{\gamma}{\epsilon} \log(\frac{1}{1 - \bigO(1)})\right)
    = \bigO \left(\frac{\gamma}{\epsilon} \right).
\end{equation}
This is precisely the scaling of the Phase Estimation algorithm.

We can reach a continuous interpolation between the classical and quantum
regimes \footnote{We refer to a ``classical'' regime whenever the quantum circuits
    involved have depth $\bigO(1)$.}\ by setting, for example,
\begin{equation}
    \eta = 1- \frac{1}{2} \left(\frac{\epsilon}{4 \gamma} \right)^{\alpha}.
\end{equation}
for $\alpha \in [0, 1]$.
From Lemma \ref{thm:projector}, the circuit depths are
\begin{align}
    D(\alpha) & = \bigO\left(\frac{\gamma}{\epsilon} \log(\frac{1}{1 -\left(\frac{\epsilon}{\gamma} \right)^{\alpha}})\right) \nonumber \\
              & = \bigO\left(\left(\frac{\gamma}{\epsilon}\right)^{1 - \alpha}\right).
\end{align}
Combining this with expression \eqref{eq:numberoftrials}, we see that the total
running time is
\begin{equation}
    \bigO\left(D(\alpha) \left(\frac{\gamma}{\epsilon}\right)^{2 \alpha}\right)
    = \bigO\left(\left(\frac{\gamma}{\epsilon}\right)^{1 + \alpha}\right).
\end{equation}

Finally, we may abandon the promise setting and convert the described procedure
into a routine that decides if the eigenenergy $\mu$ is smaller or greater than
some given threshold $\mu_0$, with tolerance for error if $\vert \mu - \mu_0
    \vert \leq \epsilon / 2$. Then, with a standard binary search scheme we solve EE
with just a $\mathrm{polylog}(\gamma / \epsilon)$ overhead to the decision
version, but keeping the same essential \aqpe\ scaling. We lay out the detailed
steps in Algorithm \ref{algo:EE}.
Applying the arguments above, we conclude the following.

\begin{theorem}
    Algorithm \ref{algo:EE} solves EE up to precision $\epsilon$ with error probability smaller than $1/3$ using a total of
    \begin{equation}
        T(\alpha) = \bigO\left(\left(\frac{\gamma}{\epsilon}\right)^{1 + \alpha} \log^2 \left( \frac{\gamma}{\epsilon} \right)\right)
    \end{equation}
    queries to $U_H$ and $U_H^{\dg}$ and with depth
    \begin{equation}
        D(\alpha) = \bigO\left(\left(\frac{\gamma}{\epsilon}\right)^{1 - \alpha}\right).
    \end{equation}
\end{theorem}

\begin{algorithm}
    \SetArgSty{}
    \caption{Eigenvalue Estimation}
    \label{algo:EE}
    \SetKwInOut{Input}{input}\SetKwInOut{Output}{output}
    \SetKwData{no}{no} \SetKwData{yes}{yes}
    \SetKwFunction{dEE}{DecideEE}
    \SetKwProg{Fn}{def}{\string:}{}
    \Input{Access to operators $U_{\psi}$, $U_H$, and $U_H^{\dg}$, the factor $\gamma$, and a precision parameter $\epsilon > 0$, as defined in Section \ref{sec:reductions}, and an interpolating parameter $\alpha$.}
    \Output{Estimate of $\mu$ up to precision $\epsilon$ with error probability smaller than $1/3$.}
    \Fn{\dEE{$\mu_0; \delta, \eta$}}{
        Block-encode $P\left(\frac{H - \mu_0 I}{\gamma + \mu_0}; \delta , \eta \right)$ as in Lemma \ref{thm:projector} and take $20 \left( \frac{4 \gamma }{ \epsilon } \right)^{2 \alpha} \lceil \log(\frac{4 \gamma }{ \epsilon}) \rceil$ \LEFT/\RIGHT\ samples\;
        Estimate probability of \RIGHT, $\overline{p}_R = (\text{number of \RIGHT\ outcomes} / \text{number of trials})$\;
        \If{$\overline{p}_R > \frac{1 - \eta + 2 \eta^2}{2}$}{
            \Return{\RIGHT}
        }
        \Else{
            \Return{\LEFT}
        }
    }
    $\delta \leftarrow \frac{\epsilon}{4 \gamma}$, $\eta \leftarrow 1- \frac{1}{2}\left(\frac{\epsilon}{4 \gamma} \right)^{\alpha}$\;
    $L \leftarrow -\gamma, R \leftarrow \gamma$\;
    \While{$ R - L > \epsilon$}{
        $\mu_0 \leftarrow \lfloor (L + U) / 2 \rfloor$\;
        \If{outcome of \dEE{$\mu_0; \delta, \eta$} is \RIGHT}{
            $L \leftarrow \mu_0$\;
        }
        \Else{
            $R \leftarrow \mu_0$\;
        }
    }
    \Return{$\mu_0$}
\end{algorithm}
}

\section{Discussion}

 {
  We have shown that the scaling of \aqpe\ can be naturally derived from the
  framework of Quantum Singular Value Transformation (at least, up to
  polylogarithmic factors in the number of samples). QSVT approaches Phase
  Estimation by block-encoding an approximation to the sign function (which is
  here adapted into a step function).  Our main contribution was to note that we
  can trade-off the quality of this approximation by longer statistical sampling.
  That is, we can compensate the use of lower degree polynomials (meaning lower
  circuit depths) by running the quantum circuits more times.

  Our derivation provides a new interpretation of the parameter $\alpha$ on \aqpe.
  Recalling that we approximate the step function up to error $\eta/2$ outside the
  interval $[-\delta, \delta]$, the approximation parameters $\eta, \delta$ and
  the interpolating parameter $\alpha$ are related as $\eta + \frac{1}{2}
      \delta^{\alpha} = 1$.  So, $\alpha$ translates how the two approximation
  parameters are related, parametrizing a family of approximations to the step
  function.

  We believe that our proof is quite intuitive, given familiarity with QSVT
  theory, as $D(\alpha)$ follows immediately from the construction of the
  polynomial approximation to the step function.  Indeed, the analysis reduces to
  checking how many samples do we have to take if we can only approximate the step
  function up to a certain degree.  This reasoning circumvents the \textit{ad hoc}
  approximations of Wang \et al.\ \cite{Wang2019}, as well as the information
  theoretic considerations of Giurgica-Tiron \et al.\ \cite{GiurgicaTiron2022}.

  When converting the decision version of Eigenvalue Estimation into the full
  search problem we gained a $\bigO(\log^2 \left( \frac{\gamma}{\epsilon}
          \right))$ factor that is not present in the original \aqpe.  Notably, this
  overhead appears only in the number of samples to collect, not in the circuit
  depth, for which our results match the state-of-the-art.  Considering the
  current landscape of noisy, small-scale quantum computing, while any overhead in
  depth can be important, such a small overhead in number of samples is arguably
  not so significant.  Regardless, it would still be interesting to find an
  optimal QSVT-based hybrid protocol.  We conjecture that by further exploiting
  the  structure of the polynomial approximation within the $[-\delta, \delta]$
  interval one could remove the polylogarithmic overhead, but a rigorous proof is
  left as an open challenge.

  Finally, our work shows that QSVT is not only a unifying framework for quantum
  algorithms, but also a useful tool to study hybrid computing.
  We suggest taking the perspective that more coherence time means better
  polynomial approximations to the target functions, reducing the need for
  repetitions.  This line of reasoning may lead to the discovery of new
  classical-quantum interpolations.
 }

\section*{Acknowledgments}

We thank Yasser Omar for introducing us to the topic of hybrid computing and
for reviewing the manuscript. We thank Diogo Cruz for reviewing the manuscript.
The authors thank the support from FCT -- Funda\c{c}\~{a}o para a Ci\^{e}ncia e
a Tecnologia (Portugal), namely through projects UIDB/50008/2020, as well as
projects QuantHEP and HQCC supported by the EU H2020 QuantERA ERA-NET Cofund in
Quantum Technologies and by FCT (QuantERA/0001/2019 and QuantERA/004/2021,
respectively). DM and MM acknowledge the support from FCT through scholarships
2020.04677.BD and 2021.05528.BD, respectively. 

\appendix

\section{Proof of Lemma \ref{thm:reductions}}
\label{ap:reductions}

{
\subsection{\texorpdfstring{$\mathrm{PE} \preceq \mathrm{AE}$}{PE reduces to AE}}

Let $(U_{\psi}, U_{\psi}^{\dg},
    \mathrm{controlled-}U,\mathrm{controlled-}U^{\dg},\epsilon )$ be an instance of
the PE problem, following the notation of Section \ref{sec:reductions}.  Let
$\ket{\psi}$ be a state of an $m$-qubit system.  Now identify the operator $A$
with the following circuit.
\[
    \Qcircuit @C=1em @R=1em {
    & \qw{^m/} & \gate{U_{\psi}} & \gate{U}  & \qw      & \qw \\
    \lstick{a\big\{ }& \qw    & \gate{H}        & \ctrl{-1} & \gate{H} & \qw
    }
\]
This makes only one call to $U_{\psi}$ and $\mathrm{controlled-}U$.
Similarly, $A^{\dg}$ only makes one call to $U_{\psi}^{\dg}$ and $\mathrm{controlled-}U^{\dg}$.
Identify $O_A$ with a $Z$ gate acting on the $a$ register.

Applying $A$ to \(\ket{0^m}\ket{0}\) yields the state (up to a global phase factor)
\begin{equation}
    A \ket{0^m} \ket{0} = \cos(\phi / 2) \ket{\psi} \ket{0} - i \sin(\phi / 2)  \ket{\psi} \ket{1}.
\end{equation}
So, we can solve AE with the instance $(A, A^{\dg}, O_A, \epsilon)$ to get an
estimate for $\phi$:  if the output of AE is $\overline{p}$, then $2
    \arccos(\overline{p})$ is an $\epsilon$-close estimate to $\phi$.  If the
algorithm for AE takes time $T$ and depth $D$, then this algorithm for PE takes
at most time $2T$ and depth $2D$.

\subsection{\texorpdfstring{$\mathrm{AE} \preceq \mathrm{EE}$}{AE reduces to EE}}

Let $(A, A^{\dg}, O_A, \epsilon)$ be an instance of the AE problem, defined on $m$ qubits, and let
\begin{equation}
    Q := A \left(2 \dyad{0^m} - I  \right) A^{\dg} O_A.
\end{equation}
Brassard \et al.\ \cite{Brassard2002} show that we can write
\begin{equation}
    A \ket{0^m} = \frac{1}{\sqrt{2}} \ket{\psi_+} + \frac{1}{\sqrt{2}} \ket{\psi_-}
\end{equation}
where $\ket{\psi_{\pm}}$ are eigenvectors of $Q$ with eigenvalues $\exp(\pm 2 i \arcsin{\sqrt{p}})$.

Identify $U_{\psi}$ with $A$, and $U_H$ with the following circuit.

\[
    \Qcircuit @C=1em @R=.5em {
    \lstick{a \big\{ } &\qw      & \gate{H} & \ctrl{1} & \ctrlo{1}      & \gate{H} \\
    &\qw{^m/} & \qw      & \gate{Q} & \gate{Q^{\dg}} & \qw
    }
    \vspace{1pt}
\]

\noindent Note that both $U_H$ and $U_H^{\dg}$ make two calls to $A$, $A^{\dg}$, and $O_A$.
$U_H$ acts as a $(1,1)$--block-encoding of the operator $\frac{Q + Q^{\dg}}{2}$, which has eigenvalue $U_{\psi} \ket{0^m}$:
\begin{align}
    \bra{0}_a U_H \ket{0}_a U_{\psi} \ket{0^m}
     & = \frac{Q + Q^{\dg}}{2} \frac{\ket{\psi_+} + \ket{\psi_-}}{\sqrt{2}}                              \\
     & = \frac{e^{2 i \arcsin{\sqrt{p}}} + e^{-2 i \arcsin{\sqrt{p}}}}{\sqrt{2}} \ket{\psi_+}  \nonumber \\
     & {} + \frac{e^{-2 i \arcsin{\sqrt{p}}} + e^{2 i \arcsin{\sqrt{p}}}}{\sqrt{2}} \ket{\psi_-}         \\
     & = (1-2 p ) U_{\psi} \ket{0^m}
\end{align}
Say we solve EE with the instance $(U_{\psi}, U_H, U_H^{\dg}, \gamma = 1, \epsilon)$, getting a value $\overline{\mu}$.
Then, $(1 - \overline{\mu}) / 2$ is an $\epsilon$-approximation to $p$.
If the algorithm for EE takes time $T$ and depth $D$, this algorithm for PE takes at most time $6T$ and depth $6D$.
}

\end{document}